\documentclass[fleqn,usenatbib,useAMS]{mnras}

\usepackage{graphicx}	
\usepackage{amsmath}
\usepackage{amssymb}	
\usepackage{bm}
\usepackage{physics} 
\usepackage{multicol}
\usepackage{tabularx}

\newcommand{\OmK}{\Omega_\text{K}}

\newcommand{\qD}{q_\mathrm{D}}
\newcommand{\qT}{q_\mathrm{T}}

\title[]{Wavelike nature of the vertical shear instability in global protoplanetary disks}

\author[Svanberg, Cui \& Latter]{
Eleonora Svanberg$^{1,2}$\thanks{E-mail: \href{mailto:eleonora.svanberg@fysik.su.se}{eleonora.svanberg@fysik.su.se}}, 
Can Cui$^{2}$ and Henrik N. Latter$^{2}$
\\
$^{1}$Department of Physics, Stockholm University, 106 91 Stockholm, Sweden \\
$^{2}$DAMTP, University of Cambridge, CMS, Wilberforce Road, Cambridge CB3 0WA, UK \\
}

\pubyear{2021}

\begin{document}
\label{firstpage}
\pagerange{\pageref{firstpage}--\pageref{lastpage}}
\maketitle

\begin{abstract}

The vertical shear instability (VSI) is a robust phenomenon in irradiated protoplanetary disks (PPDs). The majority of previous numerical simulations have focused on the turbulent properties of its saturated state. However, the saturation of the VSI manifests as large-scale coherent radially travelling inertial waves. In this paper, we study inertial-wave-disk interactions and their impact on VSI saturation. Inertial-wave linear theory is developed and applied to a representative global 2D simulation using the Athena++ code. It is found that the VSI saturates by separating the disk into several radial wave zones roughly demarcated by Lindblad resonances (turning points); this structure also manifests in a modest radial variation in the vertical turbulence strength. Future numerical work should employ large radial domains to accommodate this radial structure of the VSI, while concurrently adopting sufficiently fine resolutions to resolve the parametric instability that attacks the saturated VSI inertial waves.  

\end{abstract}


\section{Introduction}\label{in}

The vertical shear instability (VSI) is a robust phenomenon in irradiated protoplanetray disks. Originally discovered in the context of differentially rotating stars, it is related to the Goldreich--Schubert--Fricke instability \citep{gs67,fricke68} and hence is centrifugal in nature, but with double diffusive aspects \citep{bl15,lp18}. Its application to PPDs was first successfully demonstrated in \citet{nelson_etal13}. Subsequent analytical studies and numerical simulations have illustrated its linear behaviour \citep[e.g.,][]{ly15,cl21,lk22} and non-linear evolution \citep[e.g.,][]{nelson_etal13,sk14}.

The saturation of the VSI in current numerical simulations is dominated by remarkably coherent `corrugation' oscillations \citep{nelson_etal13}. However, the majority of these simulations have treated the saturation as mere turbulence and focused on including more physics \citep[e.g.,][]{flock_etal17,schafer_etal20,cb20} and exploring numerical convergence \citep[e.g.,][]{richard_etal16,manger_etal20,Flores+20}. On the other hand, linear theory reveals that (in vertically global models) the VSI is an overstability, a destabilized inertial wave \citep{bl15}; indeed, the simulated corrugation modes correspond to $n=1$, vertically standing and \textit{radially travelling} inertial waves.

Upon radial propagation, VSI corrugation modes must respond to variations in the background disk structure. For instance, 1) as they travel, their radial wavenumbers will evolve (though not their wave frequencies), possibly even reaching zero at special resonant radii (turning points); and 2) the conditions for maximum VSI growth will vary with radius for a given wave, thus an unstable travelling wave might move to a region where it does not grow as fast (or at all). It follows that the disk may divide itself into several radial wave zones. At the start of a zone, a wave is excited with the frequency that maximizes its VSI growth rate, and the end of a zone may correspond to the turning point; as it approaches its turning point, a wave is supplanted by a faster growing wave, and a new wave zone starts. It follows that each zone may possess distinct turbulence strengths, leading to the large-scale radial structuring of the disc. To accommodate the above mentioned disk-wave interactions, a wide radial simulation domain is required. Very narrow domains likely impede the radial propagation of waves, accentuate boundary effects, and thus potentially misrepresent VSI saturation. 

The paper is organised as follows. In section \ref{sec:theory}, we introduce the linear theory for inertial waves. In section \ref{sec:method}, we conduct global numerical simulation and compare the theory to the simulation data. We summarise and discuss the main findings in section \ref{sec:cd}. 

\section{Linear wave theory}\label{sec:theory}

In this section, we introduce the linear theory of inertial waves so as to facilitate the simulation data analysis in section \ref{sec:method}. The inertial waves' dispersion relation and eigenfunctions in a vertical stratified shearing box model are derived in section \ref{sec:2.1}. Their propagation over, and interactions with, a global disk are described in section \ref{sec:2.2}. The linear theory of the VSI from a purely local model is recapped in section \ref{sec:2.3} and applied to the travelling waves.

\subsection{Dispersion Relation and Eigenfunctions}\label{sec:2.1}

Working in a vertically stratified, radially localized Keplerian shearing sheet model, co-orbiting with the gas with frequency $\Omega$, the Cartesian coordinates $x,y,z$ represent the radial, azimuthal, and vertical directions, respectively. The governing equations are \citep{lp17}
\begin{align}
    &\frac{\partial \rho}{\partial t}+\mathbf{u} \cdot \nabla \rho=-\rho \nabla \cdot \mathbf{u}, \\
    &\frac{\partial \mathbf{u}}{\partial t}+\mathbf{u} \cdot \nabla \mathbf{u}=-\frac{1}{\rho} \nabla P-2 \Omega \boldsymbol{e}_{z} \times \mathbf{u}-\nabla \Phi,
\end{align}
where $\Phi = -\frac{3}{2}\Omega^2 x^2+\frac{1}{2}\Omega^2 z^2$ is the tidal potential. The gas has an isothermal equation of state, $P=c_s^{2} \rho$, where $c_s$ is a constant sound speed. 

The above equations admit the equilibrium solution 
\begin{align}
    \mathbf{u}_{0} &=-\frac{3}{2} \Omega x \boldsymbol{e}_{y}, \\
    \rho_0 &= \rho_\text{mid}\; \text{exp}[-z^2/(2H^2)], \label{4} 
\end{align}    
where $\rho_\text{mid}$ denotes the constant midplane density, and $H= c_s/\Omega$ is the pressure scale height.
This steady state is perturbed with small axisymmetric perturbations $\tilde{\rho}$, $\tilde{\mathbf{u}}$ of the form $\tilde{\rho}= \rho'(z) \mathrm{e}^{\mathrm{i} k_x x-\mathrm{i} \omega t}$, etc. We assume, without loss of generality, that $\omega>0$; thus positive $k_x$ corresponds to radially outward travelling waves, and negative $k_x$ to inward travelling waves. The linearized perturbation equations are now
\begin{align}
    &\mathrm{i} \omega u_x' = -2\Omega u_y'+\mathrm{i} k_xh', \\
    &\mathrm{i} \omega u_y' =\frac{1}{2} \Omega u_x', \\
    &\mathrm{i} \omega u_z' =\frac{d h'}{d z}, \\
    &\mathrm{i} \omega \frac{h'}{c_s^{2}} =\frac{d u_z'}{d z}+\frac{d \ln \rho_{0}}{d z} u_z'+ \mathrm{i} k_x u_x',
\end{align}
where $h'=P'/\rho_{0}$, and $P'=c_s^2\rho'$. Eliminating the velocity variables, we arrive upon a form of Hermite's equation
\begin{equation}
\frac{d^{2} h'}{d z^{2}}-\frac{z}{H^{2}} \frac{dh'}{d z}+\frac{\xi}{H^{2}} h'=0,
\end{equation}
where
\begin{equation}
\xi=\frac{\omega^{2}}{\Omega^{2}}-H^{2} k_x^{2}\left(\frac{\omega^{2}}{\omega^{2}-\Omega^{2}}\right).
\end{equation}
Solutions that have vanishing momentum as $|z|\to \infty$ require $\xi$ to be an integer $n$, and are thus Hermite polynomials of degree $n$ \citep{lp93}. The condition $\xi=n$ gives the dispersion relation
\begin{equation} \label{dispersion}
    \omega^{4}-\left[\Omega^{2}(n+1)+c_s^{2} k^{2}\right] \omega^{2}+\Omega^{4} n=0.
\end{equation}
It describes three groups of waves: the low-frequency inertial waves ($\omega<\Omega$), the high-frequency acoustic waves ($\omega>\Omega$), and density waves, for which $n=0$. 

\begin{figure}
    \centering
    \includegraphics[width=0.35\textwidth]{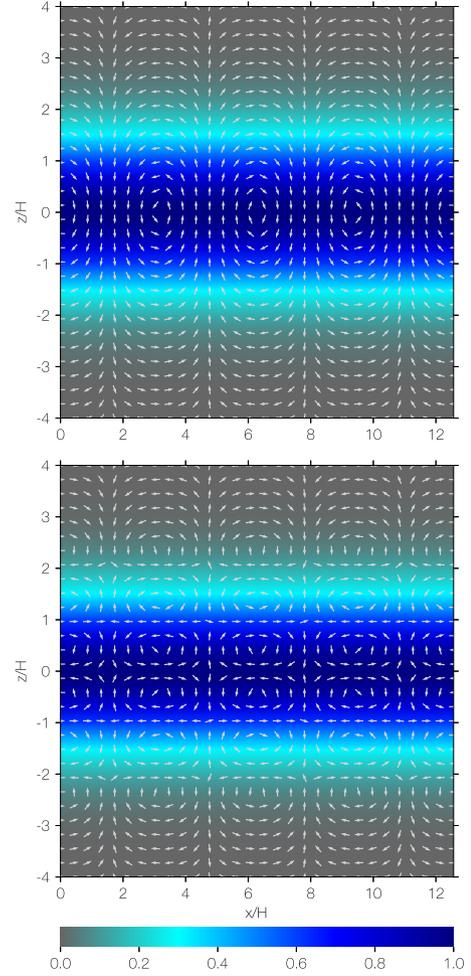} 
    \caption{Arrows: meridional velocity vectors $(u_x', u_z')$ of $n=1$ (top) and $n=6$ (bottom) inertial wave modes. Background colours: equilibrium density profile $\rho_0/\rho_\text{mid}$ by equation \eqref{4}.}
    \label{fig:ef}
\end{figure}

The waves that dominate VSI simulations are the low frequency $n=1$ inertial waves, i.e. the corrugation modes identified by \citet{nelson_etal13}. Corrugation modes possess $u_x'\propto z$ and $u_y'\propto z$, but $u_z'$ does not depend on $z$, which leads to the radial sequence of upward and downward motion familiar from simulations (all velocity components vary sinusoidally with $x$). The meridional velocity vector $(u_x', u_z')$ of a corrugation mode is plotted in the meridional plane in the top panel of Figure \ref{fig:ef}. Notice its characteristic vertical `roll' structure, which crosses the midplane. These roll features have indeed been observed in the meridional plane of recent VSI simulations \citep[see bottom panel of Figure 7 in][]{Flores+20}. In addition, we plot the higher-order $n=6$ inertial wave in the lower panel of Figure \ref{fig:ef}, which exhibits a vertical pattern of smaller rolls. These structures have only been witnessed in the highest resolution simulations of \citet{Flores+20} (top panel of their Figure 1), and in that instance have probably been excited by parametric instability attacking the primary VSI corrugation mode, as detailed in \citet{cl22}.





\subsection{Global Wave Propagation}\label{sec:2.2}

The waves described above are radially local, in that their radial wavelength is assumed to be much shorter than the length scale of any background variation in the disk. They may, however, travel over significant radial distances and thus sample the large-scale structure of the disk. As a result, their properties will evolve over radius, in particular their radial wavelength, though we expect their wave frequencies to stay fixed on the timescale of propagation  \citep{Whitham,Lighthill78}. We now describe this evolution. 

\subsubsection{Resonances or `turning points'}\label{sec:tp}

As waves travel large distances, they may encounter special radii where they interact strongly with the background flow, i.e. resonances \citep{kato2001}. The most important radii for our axisymmetric waves are where the radial wavelength approaches infinity ($k_x=0$), called turning points, which usually also correspond to Lindblad resonances.  
These turning points split the disk into radial regions where a general inertial wave can propagate and where it is prohibited. Waves that are incident on a turning point usually reflect. 
Rearranging equation \eqref{dispersion}, we obtain 
\begin{equation}
    c_s^2 k_x^2 = \frac{(\Omega^2-\omega^2)(n\Omega^2-\omega^2)}{\omega^2}.
    \label{12}
\end{equation} 
Turning points can be obtained by setting the left side equal to zero. 

Corrugation modes of $n=1$ are unique for several reasons. In a Keplerian disc, their turning points occur when $\omega=\Omega$ and correspond to a hybrid vertical/Lindblad resonance. Morover, and quite unusually, corrugation modes can propagate on either side of the turning point, and potentially can even travel through the point, despite the wave possessing an infinite wavelength there \citep[][]{bate2002}. 
If we assume a Keplerian disc and set $\Omega= \Omega_0(R/R_0)^{-3/2}$ (for fixed $\Omega_0$ and $R_0$), we find that the radial location $ R_{\rm tp}$ of the turning points are
\begin{equation}
    R_{\rm tp} = R_0 (\Omega_0/\omega)^{2/3}.
    \label{eq:tp}
\end{equation}

\subsubsection{Radial wave number}\label{sec:222}

Furthermore, as a wave packet of given frequency $\omega$ travels radially from its point of excitation, its radial wavenumber will change on account of the background disc structure. This change is described by equation \eqref{12}, which gives us $k_x$ as a function of $R$ (through $c_s$ and $\Omega$).

Far from the resonance (turning point), so that $\omega\ll\Omega$, a low-frequency $n=1$ inertial wave's wavenumber will vary as $|k_x|\approx \Omega^2/(c_s \omega)$. If, in addition, the disk possesses a background temperature structure so that $T\propto R^{-q_T}$, the radial wavenumber scales as $k_x\sim R^{q_T/2-3}$. Early VSI simulations by \citet{sk14} used $q_T=1$ and they indeed reported a radial wavenumber that varied as $R^{-5/2}$. Note that the more accurate equation \eqref{12} should be used more generally because it properly accounts for behaviour near the resonance.

\subsubsection{Phase and group velocities}\label{sec:pgv}

As discussed by \citet{lp93}, wave fronts travel radially, with the disk acting as a wave-guide. The phase speed is defined as $c_\text{p}=\omega/k_x$, which is approximately $\pm(\omega/\Omega)^2 c_s$ for $n=1$ inertial waves far from the turning point ($\omega\ll\Omega$), with the $\pm$ indicating inward or outward propagation. The radial location of any given wave crest $R_\text{p}(t)$ can then be determined from $dR_\text{p}/dt=c_\text{p}$. To facilitate comparison with our later simulations (section \ref{sec:method}), we set $q_T=1$ for the rest of this section, and thus $c_s= c_0(R/R_0)^{-1/2}$; taking the wave crests to move inwards, we find
\begin{equation}
    R_\text{p}\approx R_i\left[1+\alpha(t-t_i)\right]^{-2/3},
    \label{pv}
\end{equation}
where $\alpha= 3h\omega^2(R_i/R_0)^{3/2}/(2\Omega_0)$, $h= c_0/(\Omega R_0)$, and the crest begins at $R=R_i$ at $t=t_i$. Thus the wavecrests slow down as they move inwards.

In a vertically unstratified local model, the group velocity of a freely travelling inertial wave is perpendicular to the phase velocity. But in a vertically stratified shearing box the group and phase velocities are oppositely directed \citep{lp93}. Taking the derivative of equation \eqref{12} with respect to $k_x$, provides the following expression for the group velocity of $n=1$ inertial waves,
\begin{equation}
    c_\text{g} = \frac{d\omega}{dk_x}=\mp \frac{c_s}{1+\Omega^2/\omega^2}
    \label{15}
\end{equation}
where the minus (plus) sign is taken if the phase speed is positive (negative). Far from the resonance we have $\omega\ll\Omega$, and so $c_\text{g}\approx \mp(\omega/\Omega)^2 c_s \approx -c_\text{p}$. The group and phase velocities share the same magnitude but have opposite sign. A packet of inertial waves travels at the group speed, but wavetrains that extend over many individual wavelengths are often exposed to modulations, jumps, or other inhomogeneities in their phases and amplitudes; these `defects' also travel with speed $c_\text{g}$. If a wave defect is sufficiently localised in radius, its trajectory can be calculated by solving $dR_\text{d}/dt=c_\text{g}$, where $R_\text{d}(t)$ is the radial location of the defect. Far from resonance we can use the approximation above to obtain
\begin{equation}
    R_\text{d}(t) \approx R_i [1-\alpha (t-t_i)]^{-2/3},
    \label{eq:fit}
    \end{equation}
assuming the defect begins at $R=R_i$ at $t=t_i$ and $c_\text{g}$ points outwards. 

\subsection{Linear Growth Rate of the VSI}\label{sec:2.3}

So far we have dealt with free inertial waves, and not with the circumstances of their excitation. As discussed in early studies, the VSI provides a mechanism to drive the growth of inertial waves with certain wavevectors and frequencies \citep{nelson_etal13,bl15}. Of special interest is how the VSI chooses the frequencies of the inertial waves that ultimately dominate the disk. 
 
At a given radius we might expect that the fastest growing VSI mode will control the dynamics. To characterise this mode, we adopt the purely local incompressible model of \citet{lp18}\footnote{Ideally, we would use a vertically stratified model to determine the $k_x$ of maximum VSI growth for corrugation modes of $n=1$. Unfortunately, this is not possible as from the outset these models assume that $k_x$ is very large \citep{nelson_etal13,bl15} and prove erroneous when $k_x$ takes smaller values.}. We define the dimensionless vertical shear rate as $q_z=-R\partial\ln\Omega/\partial z$, which is of order $h$ from the thermal wind equation \citep{ks86,lk22}.
Denoting the vertical wavenumber of the mode by $k_z$, the maximum VSI growth occurs when
\begin{equation}
    \frac{k_x}{k_z} =  -\frac{1}{q_z},
\end{equation}
and growth ceases altogether for sufficiently small $|k_x/k_z|$. 
To obtain the associated wave frequency of fastest growth, we substitute $k_x$ into equation \eqref{12} and solve the quadratic equation for $\omega$. The difficulty is to relate $n$ to $k_z$. A corrugation mode has $n=1$ and thus its vertical `wavelength' will be of order the disk thickness i.e. some multiple of $H$. Anticipating our global simulation (section \ref{sec:method}), we set $k_z \sim 1/(5H)$ and $q_z \sim h = 0.05$, which gives us
\begin{equation}\label{eq:freqs}
    \omega_{\rm VSI} = (\sqrt{5}-2)\Omega.
\end{equation}
Because $\omega_{\rm VSI}$ scales as $\Omega$, the VSI favours lower frequency waves further out in the disc. Note also that $\omega_{\rm VSI}$ is always less than $\Omega$, where the resonance occurs, and thus it always lies in the inertial wave zone.

The above calculation gives a rough indication of which inertial wave frequency might be selected at any given radius. However, in reality, we do not expect to observe the $\omega$ of the waves to smoothly vary with radius. Indeed, in our simulation (Figure \ref{fig:fig9}) we see distinct wave zones emerge, each extending over a range of radii. We expect a zone that starts at a given radius to possess an $\omega$ that yields maximized VSI growth, via equation \eqref{eq:freqs}. If the wave travels outwards towards its turning point, its radial wavenumber will decrease towards 0 and its driving by the VSI will end; it may then be overtaken by a different growing VSI mode, and a new wave zone will begin.


\section{Numerical Simulations}\label{sec:method}

Global hydrodynamic simulations of the VSI are performed to verify and extend the linear inertial wave theory developed in the preceding section. We now detail our numerical methods (section \ref{3.1}), disk model (section \ref{3.2}), simulation setup (section \ref{3.3}), and simulation results (section \ref{3.4}). 

\subsection{Dynamical Equations}\label{3.1}

We employ the grid-based high-order Godunov code Athena++ \citep{stone+20} to conduct global 2D hydrodynamic simulations of the VSI. The continuity, momentum, and energy equations in their conservative form are
\begin{align}
    &\pdv{\rho}{t} + \nabla\cdot{(\rho \vb{v})} = 0, \\
    &\pdv{(\rho\vb{v})}{t}+\nabla\cdot{ ( \rho \vb{v} \vb{v}+P\vb{I} )} = - \rho\nabla{\Phi},  \\
    &\pdv{E}{t}+\nabla\cdot{(E+P)\vb{v}} =-\rho(\vb{v}\cdot \nabla \Phi) -\Lambda_\mathrm{c} \ . \label{eq:energy}
\end{align}
Here, $\vb{v}$, $\rho$, and $P$ denote gas velocity, density, and pressure, and $\vb{I}$ is the identity tensor. The total energy density is $E = \epsilon+\rho v^2/2$, where $\epsilon$ denotes the internal energy density. This is related to the gas pressure by the ideal gas equation of state $P=({\gamma-1})\epsilon$. An adiabatic index of $\gamma=7/5$ is adopted to match the molecular gas in a protoplanetary disc. The gravitational potential of the central star is given by $\Phi=-GM/r$ with stellar mass $M$, and $\Lambda_{\rm c}$ is a cooling term. 
The simulation is conducted in spherical polar coordinates $(r, \theta, \phi)$, though cylindrical coordinates $(R, z, \phi)$ are occasionally used to improve presentation.

\subsection{Disk Model}\label{sec:model}\label{3.2}

An equilibrium disk model with constant vertical temperature is adopted, i.e. the gas is vertically isothermal. The temperature and midplane density are prescribed by radial power laws,
\begin{align}
   & \rho_{\mathrm{mid}}=\rho_{0}\left(R/R_{0}\right)^{-\qD}, \\
   & c_s^2=c_0^2\left(R/R_{0}\right)^{-\qT},
\end{align}
where $\rho_0$ and $c_0$ denote density and sound speed at the inner radial boundary $R_0$.
Following \citet{nelson_etal13}, the equilibrium solutions for density and angular velocity are
\begin{align}
    & \rho(R,z) = \rho_0 (R/R_0)^{-\qD}\exp[\frac{GM}{c_s^2}\bigg( \frac{1}{r}- \frac{1}{R}\bigg)], \\
    & \Omega(R,z) = \OmK [-(\qD+\qT)h^2 + (1-\qT)+\qT R/r]^{1/2}.
\label{omega}
\end{align}
We choose $\qT=1$, and since $H \propto R^{(-\qT+3)/2}$, the aspect ratio $h=H/R$ is a constant, which we set to 0.05. 
Thus the vertical shear is
\begin{equation}
    \pdv{\Omega}{z}\approx -\frac{1}{2}\OmK \frac{z}{R^2},
\end{equation}
and hence $q_z\sim h$.
Lastly, the power-law index for density is $\qD=2$. 

Rapid Newtonian cooling is employed to excite the VSI \citep{nelson_etal13}. In the code, we adjust the temperature at each time step $\Delta t$ by 
\begin{equation}\label{temp}
    \Delta T =(T_0-T)[1-\exp(-\Delta t/\tau)].
\end{equation}
The thermal relaxation timescale $\tau$ is taken to be extremely small, so as to enforce locally isothermality; it is $10^{-10}$ innermost orbital periods.

\subsection{Simulation Setup}\label{3.3}

The radial domain of the simulation spans $r\in[1,30]$ and we use logarithmic gridding, in which the spacing increases by a constant factor of 1.005 per radial grid cell. The uniform spacing of the $\theta$ domain ranges as $\theta\in[\pi/2-5h,\pi/2+5h]$. Our 2D simulation is equipped with $704\times 448$ cells in $r \times \theta$. This allows us to reach a spatial resolution of $10$ cells per $H$ in $r$ and $12$ cells per $H$ in $\theta$.

In code units, we set $GM=R_0=1$. The simulation is run up to $3000P_0$, where $P_0=2\pi/\Omega_0$ is the orbital period at the inner boundary. To initialize the VSI, background noise is introduced to the velocities and is set to $1$ per cent of the local sound speed. Variables are set strictly to the equilibrium solution at the boundaries.

\begin{figure}
    \centering
    \includegraphics[width=0.5\textwidth]{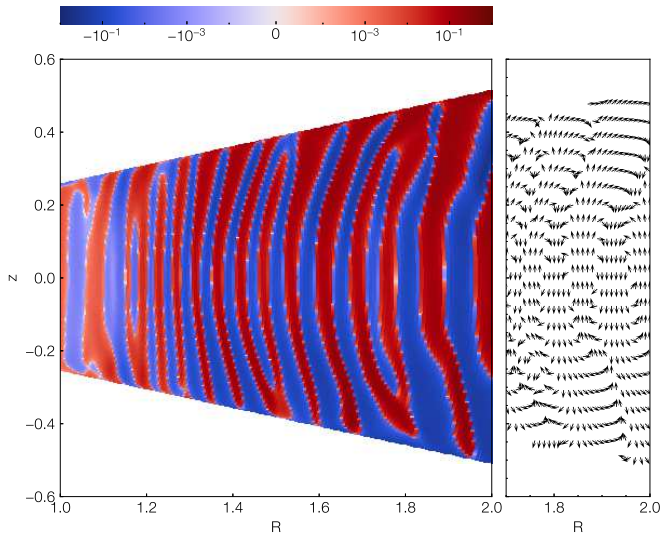}
    \caption{Left: snapshot of vertical velocity normalised by local sound speed $v_z/c_s$ at $t=500P_0$, displayed in logarithmic scale. Right: meridional velocity vector $(v_x, v_z)$ for $R\in[1.7,2]$.}
    \label{fig:vz}
\end{figure}

\begin{figure*}
    \centering
    \includegraphics[width=0.9\textwidth]{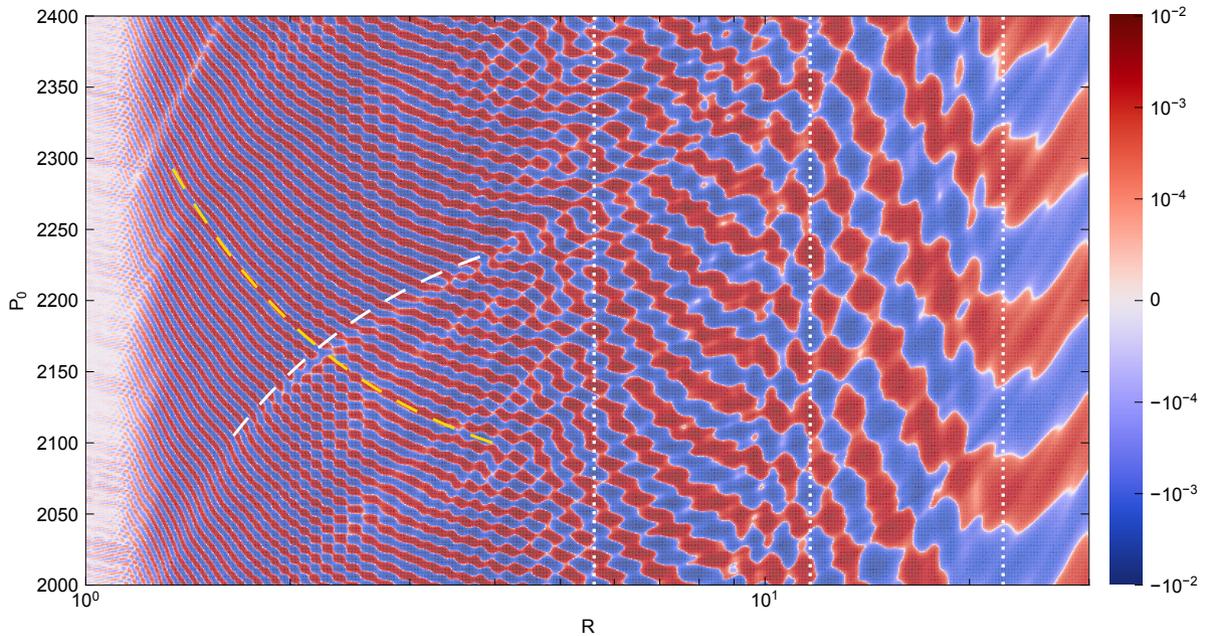}
    \caption{Space-time diagram of corrugation modes.
    Colours: strength of vertical velocity at the midplane.
    Yellow dashed curve: fitting to phase velocity with $R_i=1.65$, $t_i=2100$ by equation \eqref{pv}. 
    White dashed curve: fitting to group velocity with $R_i=3.97$, $t_i=2100$ by equation \eqref{eq:fit}. Notice that white curve is slightly shifted upwards in order to display wave defect pattern from colours.    
    Vertical dotted lines: locations of turning points (Table \ref{tab:freqrad}).}
    \label{fig:st}
\end{figure*}

\subsection{Simulation Results}\label{3.4}

\begin{figure}
    \centering
    \includegraphics[width=0.45\textwidth]{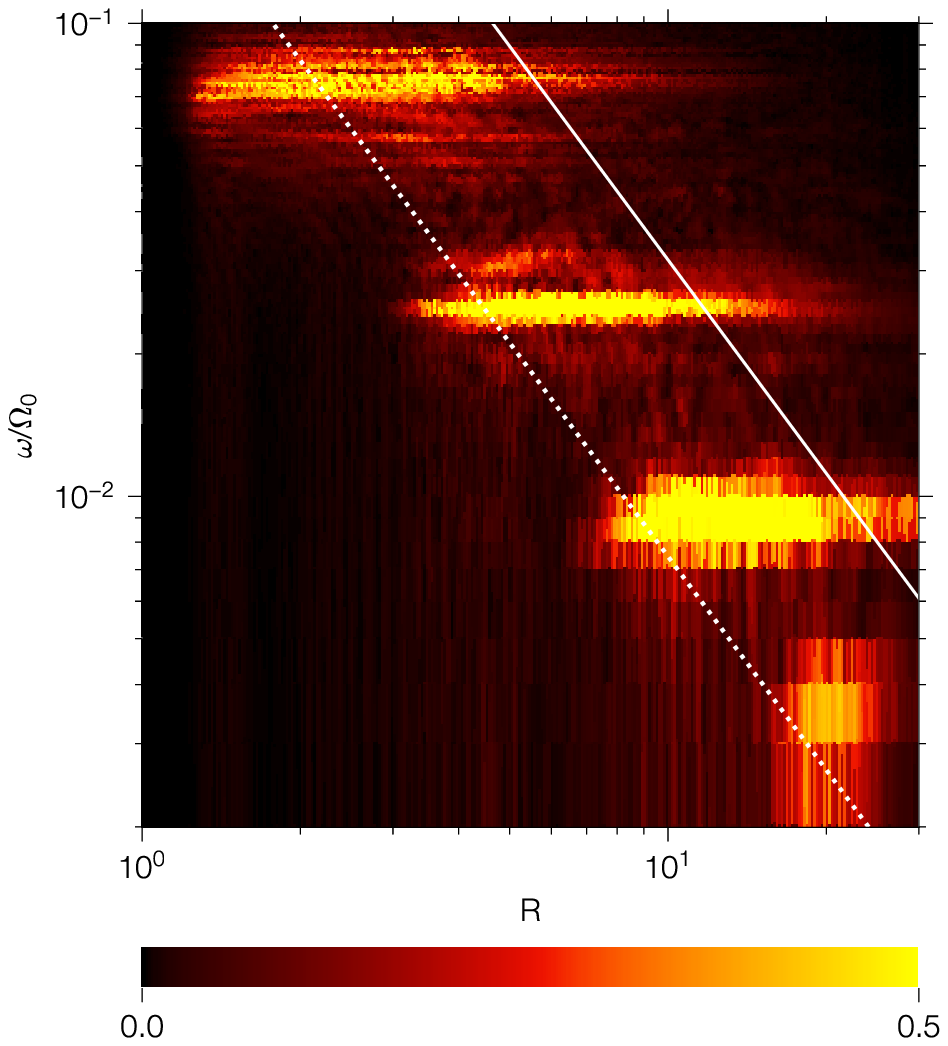}
    \caption{Unnormalized Fourier amplitudes of the corrugation modes at different radii. Dotted line denotes frequency for maximum VSI growth (equation \ref{eq:freqs}). Solid line denotes the Keplerian frequency (where resonance occurs, $\omega=\Omega$).}
    \label{fig:fig9}
\end{figure}

Here, we present our numerical simulation results and compare with the linear theory developed in section \ref{sec:theory}. Analyses are performed within a time interval from 2000 to 3000 $P_0$ and at the disk midplane.

We first give an overview of the simulation results, with a focus on wave zones, wave dynamics, and numerical boundary effects. The left panel of Figure \ref{fig:vz} shows the vertical velocity normalized by the local sound speed in the $R$-$z$ plane. The large-scale vertical oscillations witnessed here (as in all extant published VSI simulations) correspond to the inertial waves that our linear theory describes in section \ref{sec:theory}. These inertial waves possess a vertical node number $n=1$ (section \ref{sec:2.1}) because of the near absence of vertical structure and are thus corrugation modes \citep[e.g.,][]{nelson_etal13}. The right panel of Figure \ref{fig:vz} shows the meridional velocity vector $(v_x, v_z)$ but in a narrower radial domain. These circulations, obtained from the simulation, though slightly bent and possessing a steeper radial variation, have the same morphology as in the top panel of Figure \ref{fig:ef}, which shows the same quantity predicted by linear theory.

Figure \ref{fig:st} shows the space (radial) time diagram of the VSI vertical velocities. First, it is clearly seen that our simulation domain is divided into four radial wave zones, roughly demarcated by the vertical dotted lines (see below). Next, the most discernible pattern in Figure \ref{fig:st} are the coloured bands: red and blue colours correspond to the crests and troughs of radially travelling inertial waves (corrugation modes) depicted in Fig.~\ref{fig:vz}. They travel inwards and slow down as they move towards smaller radii. Narrow grey strips, especially prominent in the innermost wave zone, correspond to wave defects that propagate outwards. Furthermore, in between the wave zones, there appears an interference pattern as different waves interact. The boundaries between zones appear relatively stable over time, and thus permit a quasi-steady large-scale structure to be sustained as part of the VSI saturation. Lastly, the wave propagation is cut off by the outer radial boundary, possibly resulting in numerical artefacts in the outermost zone; our large radial domain, however, ensures the rest of the wave zone dynamics is unaffected and is physically realistic.   

\begin{figure*}
    \centering
    \includegraphics[width=0.9\textwidth]{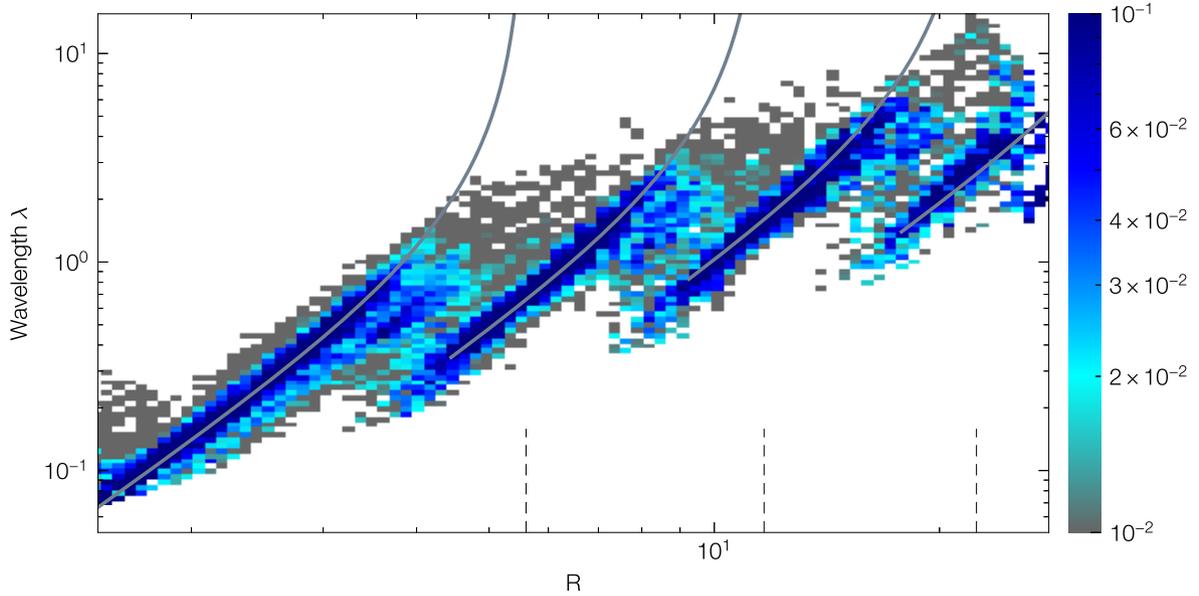}
    \caption{The radial wavelengths $\lambda$ of corrugation modes versus radius over a time interval from 2000 to 3000$P_0$. Colours, in logarithmic scale, are the probability that a specific wavelength can occur at a fixed radius. Overlaid curves are theoretical predictions by equation \eqref{12}. Vertical dashes lines are locations of turning points (Table \ref{tab:freqrad}).}
    \label{fig:fig8}
\end{figure*}

Figure \ref{fig:fig9} shows the Fourier amplitudes of the VSI corrugation modes at different radii, by performing temporal Fourier transforms of the midplane vertical velocities. It can be immediately seen that four distinct bands of frequencies appear, which correspond to the four wave zones in Fig \ref{fig:st}. We note that the frequency zones partly overlap, though Figure \ref{fig:st} indicates that normally only one wave dominates throughout its wave zone (also see Figure \ref{fig:fig8} later), presumably because it possesses the strongest Fourier amplitude. We have gone through by eye, selected one frequency associated with each band, and tabulated them in Table \ref{tab:freqrad}.

\begin{table}
 \caption{A list of frequencies and turning points.} \label{tab:freqrad}
 \begin{tabularx}{\columnwidth}{l | c c c c}
  \hline 
  & $1$ & $2$ & $3$ & $4$ \\
 \hline
  $\omega/\Omega_0 $ & $0.075$ &  $0.025$ & $0.0094$ & $0.0031$\\
  $R_{\mathrm{tp}}$ & $5.6$  & $11.7$ & $22.4$  & $47.6$\\ 
  \hline 
 \end{tabularx}
\end{table}

Now, we compare the numerical results with the analytical theory outlined in section \ref{sec:pgv} and explain features seen in Figure \ref{fig:st}. Wave crests and troughs are expected to travel at the phase speed, and hence their trajectory is delineated by equation \eqref{pv}. Indeed, the yellow dashed curve in Figure \ref{fig:st}, denoting the theoretical prediction for a selected crest, shows good agreement with the simulation data. In the theory, the phase velocity can either point radially inwards or outwards, with the corresponding group velocity possessing the same amplitude but having opposite sign (equation \ref{15}). Nevertheless, our numerical simulations, as well as \citet{sk14} and \citet{pk21}, always find the group velocity pointing outwards. One explanation is that VSI waves generated in inner regions emerge first, where the timescales are faster, and then colonise the outer regions before the VSI there has time to saturate. Finally, wave defects are expected to travel at the inertial wave group velocity. Similarly, we anticipate their trajectories to be described by equation \eqref{eq:fit}, and so overlay the theoretical prediction (white dashed curve), which perfectly agrees with the simulation data.

In Figure \ref{fig:fig9}, the solid line denotes the Keplerian frequency. The intersections between the Keplerian frequency and the yellow bands yield the locations of turning points (equation \ref{12}). These turning points are located roughly at where each wave zone terminates, consistent with Figure \ref{fig:st}. The dotted line corresponds to the frequency for the maximum VSI growth (equation \ref{eq:freqs}). The radial domain that each wave zone spans seems to be confined by these two characteristic frequencies, as anticipated in section \ref{sec:2.3}.

Figure \ref{fig:fig8} shows the radial wavelengths $\lambda$ of the corrugation modes versus radius over a time interval between 2000 and 3000$P_0$. The wavelengths are estimated by setting the distance between two consecutive sign changes in the midplane vertical velocity to equal one full wavelength \citep{sk14}. The colours, shown in logarithmic scale, represent the probability that a specific wavelength $\lambda$ occurs at a radius $R$. More specifically, at each $R$, 1000 wavelengths are obtained over the selected time interval. We bin the wavelengths, and at each $R$, calculate the probabilities by taking the ratio of data points collected in each $\lambda$ bin to the total data points collected (=1000).
It can be clearly seen that four wave zones are present. Each of them terminates near the locations of the turning points (vertical dashed lines). However, before reaching exactly the turning point, each wavetrain is supplanted by another wavetrain with frequency closer to that giving fastest VSI growth. As is quite striking, within each wave zone the wavelengths increase as the radius increase. This was indeed predicted in section \ref{sec:222}, with the result $k_x\sim R^{-5/2}$ away from the turning points. We use the more accurate equation \eqref{12} to obtain the wavelength $\lambda(R)$, with frequencies gained from Figure \ref{fig:fig9}. These predictions are overlaid in Figure \ref{fig:fig8}, which match perfectly with the coloured bands. 

\begin{figure}
    \centering
    \includegraphics[width=0.45\textwidth]{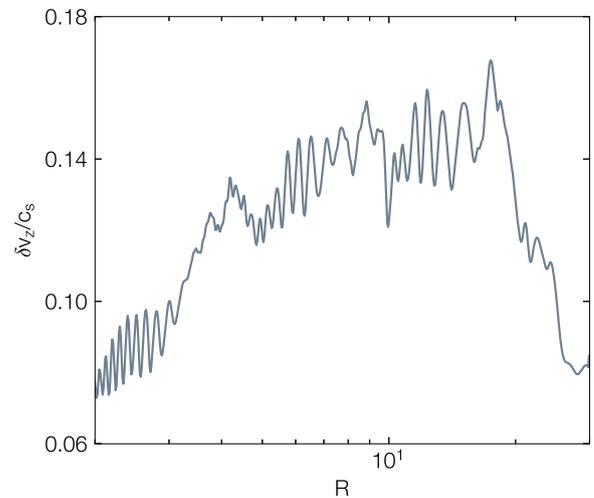}
    \caption{Level of vertical turbulence normalized by local sound speed $\delta v_z/c_s$ as a function of radius at the midplane. The equation used to calculate this quantity can be found by equation (31) in \citet{cb21}.}
    \label{fig:5}    
\end{figure}

So far, we have seen that the disk is divided into four wave zones in the simulation domain, which appear relatively stable over time (Figure \ref{fig:st}). Whether mean disk properties vary from wave zone to wave zone is thereby of interest. Figure \ref{fig:5} shows the root mean square vertical velocity $\delta v_z$ normalized by the local sound speed. It is clearly seen that the mean vertical turbulence level increases with radius. Superimposed on this mean increase is a sawtooth pattern, separated into four zones, roughly coincident with the radial turning points. The level of variation in $\delta v_z/c_s$ is about a factor of two between the highest and lowest values in our domain. Therefore, a modest correlation between the saturated vertical level of turbulence and wave zones is found. However, other disk properties are found to be very weakly affected, consistent with previous numerical simulations with extended radial domains \citep[e.g.,][]{flock_etal17,flock_etal20,cb20}.

\section{Conclusions and Discussion}\label{sec:cd}

In this paper, we revisited the global VSI hydrodynamic simulation and studied its wavelike properties upon saturation. As witnessed in previous simulations, the non-linear saturation of the VSI is dominated by radially travelling corrugation modes ($n=1$ inertial waves). In the first part of this paper, a linear theory for these waves is developed in a vertically global but radially local model, and their response to the background disk structure upon radial propagation over a global disk is determined. In the second part of this paper, a representative 2D hydrodynamic simulation was conducted with the Athena++ code and compared with the linear theory. 

The comparison between the theory and the simulation results is satisfactory. It is found that the simulation domain is separated into four radial wave zones. Each zone roughly starts at locations that maximize the linear VSI growth rate and ends near the locations of the turning points (resonances), where the modes' wavelengths go to infinity. In each wave zone, inertial wave crests travel radially inwards at the phase velocity. The wave defects travel radially outwards at the group velocity. Wave zones possess vertical turbulence levels that are modestly distinct from each other. 
 
In future work, it is necessary for numerical simulations to employ large radial domain to accommodate the VSI wave propagation, in particular to describe distinct wave zones. 
This paper studies the large-scale coherent corrugation modes that dominate in low resolution global simulations. This coherence could, however, be disrupted by parametric instability, though this requires finer resolutions to resolve \citep{cl22}. It should be a high priority to verify the saturation properties are qualitatively correct when the parametric instability presents.  

\section*{Acknowledgements}

We thank the anonymous referee, whose comments helped improve and clarify this manuscript, and Gordon Ogilvie who pointed out some errors in a previous version of the manuscript.
We thank Lizxandra Flores-Rivera for useful discussions. ES acknowledges the support from Philippa Fawcett Internship Programme at Centre for Mathematical Sciences, University of Cambridge. CC and HNL acknowledge funding from STFC grant ST/T00049X/1. Numerical simulations are conducted on the FAWCETT cluster at the Department of Applied Mathematics and Theoretical Physics, University of Cambridge.

\section*{Data Availability}

The data underlying this article will be shared on reasonable request to the corresponding author.


\appendix


\bibliographystyle{mnras}
\bibliography{disk} 


\bsp	
\label{lastpage}
\end{document}